\documentclass[runningheads]{llncs}

\usepackage{amsmath}
\usepackage{graphicx}
\usepackage{url}

\usepackage{booktabs}
\usepackage{xcolor}
\usepackage{textcomp}
\usepackage{tikz}
\usetikzlibrary{positioning, arrows.meta, fit, backgrounds}
\usepackage[hidelinks,breaklinks=true]{hyperref}

\title{Persona-as-Configuration: Generative Stakeholder Reporting for Agricultural Floods}

\titlerunning{Persona-as-Configuration for Agricultural Flood Reporting}

\author{Oliver Aleksander Larsen \and
Tiziano Santilli \and
Francesco Daghero \and
Mahyar T. Moghaddam}

\authorrunning{O. A. Larsen et al.}

\institute{University of Southern Denmark, Maersk Mc-Kinney Moller Institute,\\
Odense, Denmark\\
\email{\{olar,tisa,fdag,mtmo\}@mmmi.sdu.dk}}

\begin{document}
\maketitle
{\let\thefootnote\relax\footnotetext{This version of the contribution has been accepted for publication after peer review, but is not the Version of Record and does not reflect post-acceptance improvements or corrections. Accepted at CASA 2026 (9th Workshop on Context-Aware, Autonomous and Smart Architectures), co-located with ECSA 2026. The Version of Record will appear in the Springer LNCS post-proceedings of ECSA 2026.}}

\begin{abstract}
Cyber-physical systems built on deterministic edge inference, such as on-vehicle flood detection for agricultural fields, produce structured decision logs that must be interpreted differently by heterogeneous stakeholders. Pairing such systems with large language models (LLMs) to generate stakeholder-specific reports introduces a tension: the generative layer is non-deterministic, while the edge plane must remain replayable and auditable. We propose an architectural pattern resting on two invariants: \emph{unidirectional consumption}, in which the generative layer is a strict read-only consumer of the deterministic plane and never writes back, and \emph{persona-as-configuration}, in which stakeholder adaptation is a versioned prompt-template artifact rather than runtime improvisation. We instantiate the pattern as a context-aware dashboard layer over the JSON decision logs of a previously published edge-based standing-water detection system, and analyse how the integration boundary admits standard generative-reliability mitigations as configuration- or middleware-level extension points. A structured expert review rated the pattern favourably across five ISO/IEC~25010-aligned quality dimensions, with strongest agreement on separation of concerns. End-user evaluation with agricultural stakeholders is planned for future work.

\end{abstract}

\section{Introduction}
\label{sec:introduction}

Standing water in agricultural fields threatens vehicle mobility, crop viability, and soil structure, especially on narrow off-road tracks where drivers must make rapid go/no-go decisions under changing environmental conditions~\cite{edge_agriculture}.
Detecting such hazards in real time requires inference on commodity edge hardware operating under intermittent connectivity and constrained power budgets~\cite{iot_edge_rfc}.
Yet accurate detection alone is insufficient: the value of a flood-monitoring system depends on whether the right people can act on its outputs.

Six stakeholders need distinct outputs: farmers (field accessibility), agronomists (soil saturation), farm managers (route safety), government agencies (audit-ready documentation), insurers (georeferenced damage evidence), and decision-makers (cross-role executive summary).
Translating a single set of sensor readings into these varied narratives requires natural-language generation that is both data-grounded and role-aware.
Large language models (LLMs) offer this capability, yet integrating a non-deterministic generative component into a deterministic edge system raises architectural questions that have received little attention.

The LLM in this architecture is not a report generator bolted onto a monitoring system: it is the \emph{mechanism} by which a single deterministic event stream is multiplexed into $N$ stakeholder-appropriate views.
Without it, persona-based reporting collapses into $N$ hand-written templates that are infeasible at the cardinality of sensor-pattern $\times$ anomaly-type $\times$ persona and cannot synthesise spatial clusters or correlate anomaly magnitudes. The persona-as-configuration architectural element ceases to exist.
The architectural function of the generative component is therefore multiplexing-via-personas, not summarisation, positioning this work within the family of context-aware, smart cyber-physical architectures.

We propose a two-layer architecture for agricultural flood monitoring that addresses both detection and stakeholder communication.
The \emph{edge detection layer}, contributed in our prior work~\cite{larsen_edge_detection}, classifies standing-water hazards on resource-constrained on-vehicle hardware and persists structured decision logs. This paper treats it as a black-box producer.
The \emph{human-interface layer} consumes those logs through a web-based dashboard and generates stakeholder-tailored reports using a cloud-hosted LLM, and is the locus of the two contributions that follow.

\textbf{C1 (Unidirectional consumption invariant).} A strict read-only-consumer relationship from the generative layer to the deterministic edge plane admits LLM non-determinism into a safety-relevant cyber-physical system without compromising replayability or audit.

\textbf{C2 (Persona-as-configuration pattern).} Stakeholder adaptation is elevated to a first-class configuration element by mapping each stakeholder type to a versioned prompt template, applying the \emph{Persona} pattern~\cite{white_prompt_patterns} at the architectural rather than the prompt-engineering level.

Both contributions are assessed through a structured expert review tied to the ISO/IEC~25010 product quality model~\cite{iso_25010}. Empirical end-user evaluation is planned for future work.

\section{Background and Related Work}
\label{sec:background}

\subsection{Standing-Water Detection and Edge Perception}

Detecting standing water for off-road vehicles spans satellite mapping, fixed camera networks, and on-vehicle edge perception~\cite{flood_sensors_review}.
Our prior work~\cite{larsen_edge_detection} contributes a deployed edge-based detection system that uses finite-state-machine-guided model tiering and multi-model consensus on Raspberry-Pi-class hardware. The present paper treats that system as a black-box producer of JSON decision logs and focuses entirely on the human-interface layer that consumes them.

\subsection{LLMs for Operational Decision Support}

Generating stakeholder-appropriate text from structured data has been studied as data-to-text natural language generation for over two decades~\cite{reiter_dale_nlg}. Large language models extend this work to less templated, narrative outputs and are increasingly applied to structured-data interpretation for non-technical stakeholders.
An et al.\ proposed IoT-LLM, a framework that preprocesses Internet of Things (IoT) signals into LLM-amenable formats with retrieval-augmented generation (RAG~\cite{gao_rag_survey}) and chain-of-thought reasoning~\cite{iot_llm}.
A systematic review of LLMs in IoT identifies data interpretation and contextual reasoning as the principal gains over rule-based analytics~\cite{llm_iot_survey}.
In agriculture, IoT and AI integration have reshaped precision farming~\cite{iot_ai_agriculture,cloud_edge_agriculture}, yet the focus remains on sensing rather than generating stakeholder-specific narratives from sensor outputs.
A review of LLM applications in disaster management identifies insufficient integration across stakeholder groups and inadequate transformation of situational data into actionable insights as critical gaps~\cite{llm_disaster_review}. Our architecture addresses both by generating stakeholder-differentiated reports from edge decision logs.

Beyond model selection, prompt engineering plays a key role in tailoring LLM outputs to specific audiences.
White et al.\ introduced a prompt pattern catalog including the \emph{Persona} pattern~\cite{white_prompt_patterns}. Schreiber et al.\ later extended this catalog into a pattern language with multi-persona and dynamic-switching variants at the prompt-engineering level~\cite{schreiber_persona_language}. Schulhoff et al.\ presented a taxonomy of 58~prompting techniques~\cite{schulhoff_prompt_report}.
A survey on persona in LLMs distinguishes \emph{role-playing} (assigning a persona to the model) from \emph{personalization} (adapting to a user's persona)~\cite{persona_survey}, and recent empirical work shows that role assignment alone in system prompts does not consistently improve LLM task performance~\cite{zheng_helpful_assistant}.
Our claim is correspondingly architectural: we apply the Persona pattern as a first-class architectural element, mapping each stakeholder type to a versioned prompt configuration that treats role-based prompting as a unit of governance, audit, and substitution, not as an accuracy lever.

\subsection{Dashboard Architectures for IoT and Flood Monitoring}

IoT application architectures route sensor networks through layered topologies~\cite{iot_dashboards_survey} to human-facing dashboards providing visualization, alerting, and trend analysis.
FAIS combines IoT APIs, USGS camera feeds, and convolutional neural network classifiers into a Python web dashboard with map-based flood analytics~\cite{fais}.
Recent systems integrate ultrasonic and pressure sensors with web dashboards and Firebase alerting for real-time flood-risk classification~\cite{flood_dashboard_iot}.
These architectures are typically cloud-centric (sensors stream data to a central server that renders dashboards) and focus on visualization and alerting rather than natural-language interpretation.

Read-only consumption downstream of the sensing tier is itself standard in this IoT application layer~\cite{iot_dashboards_survey}; what changes the problem here is a non-deterministic consumer. Our architecture consumes offline JSON artifacts from a resource-constrained edge system through a generative module, so the Layer~1 to Layer~2 boundary becomes a deliberately stated invariant that protects the deterministic edge plane, not an incidental flow direction, and is paired with persona-as-configuration and the reliability-seam analysis of \S\ref{sec:gen_reliability}.

\subsection{Architectural Patterns for AI-Integrated Systems}

Software architecture research has recently begun to address LLM integration challenges, extending the ML-systems architectural agenda~\cite{sculley_hidden_debt} with pattern-oriented techniques~\cite{bass_software_architecture}.
Bucaioni et al.~\cite{bucaioni_ref_arch} proposed a functional reference architecture for LLM-integrated systems, noting the absence of commonly accepted architectural blueprints.
Shamsujjoha et al.~\cite{swiss_cheese_ai} adapted the Swiss Cheese Model into multi-layered runtime guardrails for agents based on foundation models, with each layer independently protecting specific quality attributes.
Xu et al.~\cite{llm_cps_survey} surveyed LLM-enabled cyber-physical systems (CPS) and categorised the roles LLMs play, highlighting the tension between probabilistic LLM reasoning and deterministic CPS requirements.

Our architecture resolves this tension through strict unidirectional data flow, not runtime guardrails.
Where Bucaioni et al.~\cite{bucaioni_ref_arch} catalogue functional blocks and Shamsujjoha et al.~\cite{swiss_cheese_ai} layer runtime guardrails, we make the boundary itself a load-bearing architectural invariant (\S\ref{sec:two_layer}). Closest in spirit, Gonz\'alez-Potes et al.~\cite{gonzalez_hybrid_llm} pair deterministic rule-based supervision with an LLM advisory channel for industrial batch control. The safety partition is shared, but not the stakeholder-identity-as-configuration treatment of \S\ref{sec:reporting}. The cited guardrails would attach at the seams identified in \S\ref{sec:gen_reliability}.

\section{Architecture}
\label{sec:architecture}

\begin{figure}
\centering
\resizebox{0.92\columnwidth}{!}{%
\begin{tikzpicture}[
    node distance=0.45cm and 0.5cm,
    box/.style={draw, rounded corners, minimum height=0.75cm, minimum width=2.0cm, font=\scriptsize, align=center},
    edgebox/.style={box, fill=blue!12, minimum width=4.5cm, minimum height=1.0cm},
    dashbox/.style={box, fill=green!12},
    llmbox/.style={box, fill=red!12},
    jsonbox/.style={box, fill=orange!15, minimum width=3.4cm},
    outbox/.style={box, fill=yellow!20, minimum width=2.0cm},
    arr/.style={-{Stealth[length=2.5mm]}, thick},
    darr/.style={-{Stealth[length=2.5mm]}, thick, dashed},
    invarr/.style={{Stealth[length=2.0mm]}-, thick, red!70, dotted},
    lbl/.style={font=\tiny, midway},
]
\node[edgebox] (edge) {Edge Detection System\\\scriptsize (FSM tiering, multi-model consensus,\\sensor fusion)~\cite{larsen_edge_detection}};

\begin{scope}[on background layer]
    \node[draw=blue!50, fill=blue!4, rounded corners, dashed,
          fit=(edge),
          inner sep=6pt,
          label={[font=\scriptsize\bfseries, blue!70]above:Layer 1: Edge Detection (on-vehicle)}] (l1) {};
\end{scope}

\node[jsonbox, below=0.85cm of edge] (json) {JSON Decision Logs\\\scriptsize (Layer~1 / Layer~2 contract)};
\draw[arr] (edge.south) -- node[lbl, right] {writes} (json.north);

\node[dashbox, below=0.85cm of json, xshift=-2.0cm] (dl) {data\_loader};
\node[dashbox, right=0.45cm of dl] (mg) {map\_generator};
\node[dashbox, right=0.45cm of mg] (db) {dashboard};
\node[llmbox, below=0.55cm of db] (llm) {llm\_report};
\node[outbox, left=0.5cm of llm] (rep) {Report (.txt)};

\draw[arr] (json.south) -- ++(0,-0.25) -| (dl.north);
\draw[arr] (dl) -- (mg);
\draw[arr] (mg) -- (db);
\draw[darr] (db.south) -- node[lbl, right, xshift=1pt] {on-demand} (llm.north);
\draw[arr] (llm) -- (rep);

\begin{scope}[on background layer]
    \node[draw=green!50!black, fill=green!4, rounded corners, dashed,
          fit=(dl)(mg)(db)(llm)(rep),
          inner sep=6pt,
          label={[font=\scriptsize\bfseries, green!50!black]below:Layer 2: Human Interface (base station)}] (l2) {};
\end{scope}

\draw[invarr] ([xshift=4pt]edge.east) to[bend left=35] ([xshift=4pt]json.east);
\node[font=\tiny\itshape, red!70, anchor=west, text width=2.6cm, align=left]
  at ([xshift=0.45cm,yshift=-0.05cm]json.east) {Unidirectional: Layer~2 never writes back};

\begin{scope}
  \node[draw, rounded corners, fill=gray!5, inner sep=4pt, font=\tiny, align=left,
        anchor=north west]
    at ([xshift=0.5cm]l2.north east) {%
      \begin{tabular}{@{}c@{~~}l@{}}
        \tikz\node[edgebox, minimum width=0.4cm, minimum height=0.25cm, font=\tiny]{}; & Edge / deterministic \\
        \tikz\node[dashbox, minimum width=0.4cm, minimum height=0.25cm, font=\tiny]{}; & Human interface / advisory \\
        \tikz\node[llmbox, minimum width=0.4cm, minimum height=0.25cm, font=\tiny]{}; & Generative AI boundary \\
        \tikz\node[jsonbox, minimum width=0.4cm, minimum height=0.25cm, font=\tiny]{}; & JSON contract artifact \\
        \tikz\node[outbox, minimum width=0.4cm, minimum height=0.25cm, font=\tiny]{}; & Generated report (output) \\
        {\color{red!70}\rule[2pt]{0.4cm}{0.4pt}} & No-feedback invariant \\
      \end{tabular}
    };
\end{scope}
\end{tikzpicture}}%
\caption{Two-layer architecture. Layer~1 (edge detection, described in~\cite{larsen_edge_detection}) writes JSON decision logs that fix the Layer~1 / Layer~2 contract. Layer~2 (this paper's contribution) consumes them for visualization and on-demand stakeholder-tailored LLM reporting. Data flows strictly downward. The unidirectional invariant prevents the generative layer from writing back to the safety-critical detection plane.}
\label{fig:architecture}
\end{figure}
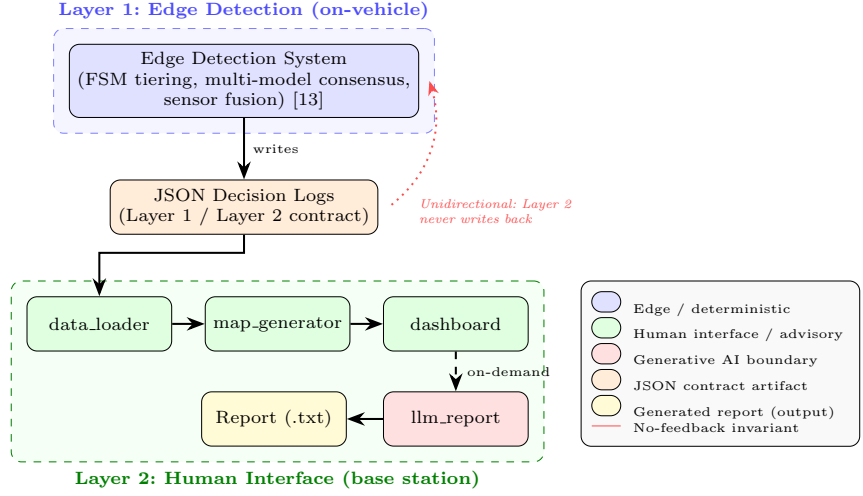

This section presents the system architecture in three parts: the architectural drivers that shape the design, the two-layer organisation and its key invariants, and the dashboard component decomposition with the LLM integration point characterised as an architectural boundary.
Figure~\ref{fig:architecture} provides an overview.

\subsection{Architectural Drivers}
\label{sec:drivers}

Six drivers shape the architecture and are annotated downstream where each is satisfied.
\textbf{D-Det} (Determinism of the safety-critical path): the edge plane must be replayable bit-for-bit independent of the generative layer, driving the unidirectional invariant of \S\ref{sec:two_layer}.
\textbf{D-Rep} (Replayability and audit): every report must be reconstructable from the underlying decision log, driving the schema-stable JSON contract of Table~\ref{tab:json_fields}.
\textbf{D-Stk} (Stakeholder heterogeneity): one deterministic stream must be rendered as $N$ role-appropriate views, driving persona-as-configuration in \S\ref{sec:reporting}.
\textbf{D-Cost} (Bounded LLM cost): a cloud LLM call is orders of magnitude more expensive than a per-frame edge inference and cannot be invoked per frame, driving on-demand invocation in \S\ref{sec:llm_boundary}.
\textbf{D-Conn} (Intermittent agricultural connectivity): the edge plane runs offline, so cloud-bound interpretation must be deferrable through batch ingest of JSON at the base station.
\textbf{D-Sub} (Independent provider evolution): the chosen LLM may need to be swapped (e.g.\ cloud-to-local) without touching surrounding code, driving the substitutable \texttt{llm\_report} interface of \S\ref{sec:llm_boundary}.

The relevant reference frame for these drivers is not application architecture (hexagonal, clean, MVC) but cyber-physical control-plane separation~\cite{llm_cps_survey,swiss_cheese_ai}: the dominant concern is the safety boundary between deterministic and non-deterministic compute, not the inversion of business-logic dependencies. A hexagonal decomposition could be applied within Layer~2 (data-loader, presenter, LLM-adapter as ports) but would not address the cross-layer invariant that defines the architecture and that D-Det and D-Rep require.

\subsection{Two-Layer Organisation and the Unidirectional Invariant}
\label{sec:two_layer}

\textbf{Layer~1 (Edge Detection).}
The edge layer, described in detail in our prior work~\cite{larsen_edge_detection}, runs on-board a vehicle traversing agricultural fields.
It combines a Raspberry-Pi-class capture node, a processing node that owns a finite-state machine for adaptive model tiering, and an optional NVIDIA Jetson worker for GPU-accelerated inference. Environmental sensor anomalies are fused with multi-model vision consensus to classify each frame as 0~(No~Flood), 1~(Some~Water), or 2~(Flooded).
What matters here is the contract it exposes.
The Layer~1 / Layer~2 contract is a per-sensor JSON record (Table~\ref{tab:json_fields}), and every design choice that follows is made against this contract rather than against the implementation that produces it.

\textbf{Layer~2 (Human Interface)} runs on a base-station workstation, ingests Layer~1's JSON artifacts, and provides two capabilities: (a)~an interactive colour-coded flood map rendering the geo-referenced snapshot on a Leaflet map via Folium~\cite{folium}, and (b)~on-demand stakeholder-tailored natural-language reports generated by GPT-5-mini~\cite{openai_gpt5mini}.

The central design invariant is that data flows \emph{strictly from Layer~1 to Layer~2}.
This guarantees the safety-critical detection path remains unaffected by the non-deterministic generative component.
The LLM is advisory: it interprets decisions already made, rather than influencing future ones.

\begin{table}
\centering
\caption{Key per-sensor JSON record fields consumed by Layer~2 (a timestamp and detection-image are also present in the schema).}
\label{tab:json_fields}
\begin{tabular}{@{}ll@{}}
\toprule
Field & Description \\
\midrule
\texttt{camera\_id}        & Sensor / camera identifier \\
\texttt{location}          & GPS coordinates \texttt{[lat, lon]} \\
\texttt{sensor\_data}      & Temperature (\textdegree C), humidity (\%), pressure (hPa) \\
\texttt{sensor\_anomalies} & $\Delta T$, $\Delta\mathrm{RH}$, $\Delta P$ vs.\ diurnal baseline \\
\texttt{prediction}        & Hazard classification: 0 / 1 / 2 \\
\texttt{scores.combined}   & Weighted flood score. Higher means stronger evidence \\
\bottomrule
\end{tabular}
\end{table}

\subsection{Dashboard Component Decomposition}

\begin{figure}
\centering
\includegraphics[width=\textwidth]{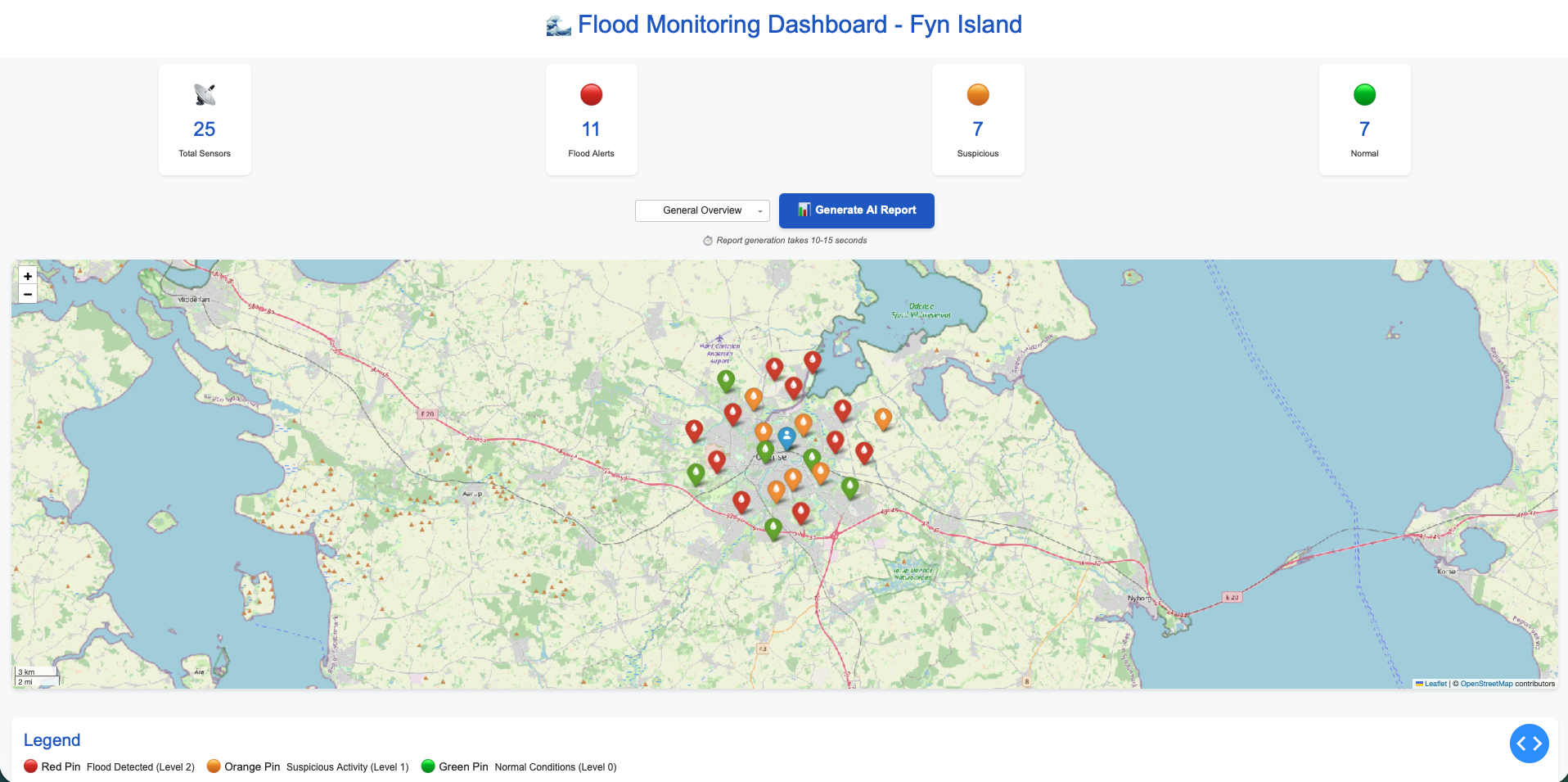}
\caption{Dashboard prototype showing the Fyn Island monitoring area. Colour-coded map pins indicate per-sensor flood classifications (red = Flooded, orange = Suspicious, green = Normal). Summary cards report aggregate counts.  The \emph{Generate AI Report} button triggers stakeholder-tailored report generation via GPT-5-mini.}
\label{fig:dashboard}
\end{figure}

Layer~2 is implemented as a modular Python application with five modules (Table~\ref{tab:modules}). Figure~\ref{fig:dashboard} shows the running prototype.
At runtime, \texttt{data\_loader} reads JSON files into sensor-record dictionaries. \texttt{map\_generator} renders them as colour-coded Folium markers. \texttt{dashboard} embeds the map, computes summary counts, and renders the UI.
When the operator selects a stakeholder type and clicks \emph{Generate AI Report}, \texttt{llm\_report} constructs a prompt (\S\ref{sec:reporting}), calls GPT-5-mini, displays the result, and writes a timestamped copy.

\begin{table}
\centering
\caption{Module decomposition of the human-interface layer.}
\label{tab:modules}
\begin{tabular}{@{}lp{7cm}@{}}
\toprule
Module & Responsibility \\
\midrule
\texttt{config}          & Deployment constants: colour-code mapping, data paths, classification thresholds. \\[1pt]
\texttt{data\_loader}    & Parses JSON sensor files from \texttt{data/} and aggregates readings into an in-memory list of sensor records. \\[1pt]
\texttt{map\_generator}  & Produces an interactive Folium~\cite{folium} map with colour-coded markers (red = Flooded, orange = Suspicious, green = Normal) and click popups showing sensor data and detection images. \\[1pt]
\texttt{dashboard}       & Hosts the Dash~\cite{dash} web application: embeds the Folium map, renders summary cards, provides a stakeholder dropdown, and exposes a \emph{Generate AI Report} callback. \\[1pt]
\texttt{llm\_report}     & Generative-AI boundary (\S\ref{sec:llm_boundary}): accepts sensor data + stakeholder type, constructs a role-specific prompt, calls GPT-5-mini, returns and persists report text. \\
\bottomrule
\end{tabular}
\end{table}

\subsection{LLM Integration as an Architectural Boundary}
\label{sec:llm_boundary}

The \texttt{llm\_report} module is the controlled entry point through which the generative component interacts with the system.
The boundary is designed around four properties tied to the architectural drivers of \S\ref{sec:drivers}.
\emph{On-demand invocation} (D-Cost) calls the LLM only when an operator explicitly requests a report, keeping the dashboard responsive and bounding API spend.
\emph{Statelessness} packages each report fresh from the JSON record and stakeholder type, with no conversation history across calls. This prevents context-window drift and makes the prompt bit-exact reconstructible from the JSON record and stakeholder type.
\emph{Auditability} (D-Rep) pairs every generated report with a timestamped copy alongside the source JSON record and the version-controlled prompt template, so the chain from edge decision to stakeholder report remains reconstructable despite LLM phrasing variability.
\emph{Substitutability} (D-Sub) confines provider choice to a narrow interface (sensor data and stakeholder type in, report text out), so GPT-5-mini can be replaced by a different cloud provider or by a local model without touching dashboard code.

\section{Stakeholder-Aware Report Generation}
\label{sec:reporting}

\subsection{Prompt Configuration Architecture}

Each of the six supported stakeholder types (Table~\ref{tab:stakeholders}) maps to a structured prompt configuration with three elements:
{\em 1)}~a \emph{system prompt} assigning a domain-specific persona and report structure;
{\em 2)}~a \emph{data injection template} serialising aggregated sensor records; and
{\em 3)}~a \emph{focus directive} enumerating priority topics.

\begin{table}
\centering
\caption{Stakeholder prompt configurations and report focus areas.}
\label{tab:stakeholders}
\begin{tabular}{@{}lp{7cm}@{}}
\toprule
Stakeholder & Report Focus \\
\midrule
General Overview     & Executive summary, overall risk level, prioritised recommendations. \\[1pt]
Farmer               & Field accessibility, crop damage potential, livestock safety, immediate action items. \\[1pt]
Agronomist           & Soil health indicators, cross-sensor anomaly patterns, drainage and remediation advice. \\[1pt]
Farm Manager         & Vehicle and route safety, detection confidence trends, workforce scheduling. \\[1pt]
Government Agency    & Regulatory compliance, audit-ready documentation, inter-agency coordination. \\[1pt]
Insurance Company    & Geo-referenced damage evidence, quantified loss estimates, temporal claims records. \\
\bottomrule
\end{tabular}
\end{table}

This design applies the Persona pattern~\cite{white_prompt_patterns} at the architectural level: each persona is a version-controlled configuration artifact, not an ad-hoc instruction.
The architectural consequence is concrete: a new stakeholder type should be addable by a non-prompt-engineer (a domain expert) because the artifact has a stable schema and is reviewable independently of the data pipeline, dashboard, or LLM integration code. Nothing else is touched.

\subsection{Data-to-Prompt Pipeline}

When a report is requested, the module executes four steps.

\textbf{1.~Aggregation.}
All sensor records are serialised into a structured text block.
Each record is rendered as a compact summary: timestamp, sensor identifier, GPS coordinates, environmental readings, anomaly deltas ($\Delta T$, $\Delta\mathrm{RH}$, $\Delta P$), classification label, and combined score.
Including the full sensor network lets the LLM synthesise spatial patterns such as clusters of flood alerts.

\textbf{2.~Template composition.}
The data block is inserted into the selected stakeholder's prompt template, which specifies the persona (e.g., ``You are an agronomist advising on post-flood soil remediation''), expected sections (risk assessment, affected areas, recommendations), and focus directive from Table~\ref{tab:stakeholders}.

\textbf{3.~API submission.}
The composed prompt is submitted to GPT-5-mini as a single-turn request: a system message (persona and instructions) and a user message (sensor data and generation directive).
No multi-turn conversation is maintained, enforcing the statelessness property described in \S\ref{sec:llm_boundary}.

\textbf{4.~Persistence and presentation.}
The returned text is displayed in a modal overlay and simultaneously written to a timestamped \texttt{.txt} file for auditing.

\subsection{Illustrative Output}

Table~\ref{tab:report_excerpts} shows excerpts from the six persona reports generated from one sensor record: the JSON data block in the user message is held fixed, and only the prompt-template configuration (the system-message persona and focus directive) changes.
For contrast, the farmer receives plain-language accessibility guidance while the insurer receives claims-oriented damage evidence, showing how the Persona configuration produces operationally distinct outputs without any change to the data pipeline.

\begin{table}
\centering
\caption{Excerpts generated from the same 25-sensor decision-log batch (Fyn Island, 2026-02-15; 11 stations at Level~2, score range 0.080 to 3.003). All six stakeholder personas are produced from one JSON input. Only the prompt-template configuration changes.}
\label{tab:report_excerpts}
\begin{tabular}{@{}lp{7.7cm}@{}}
\toprule
Stakeholder & Generated Excerpt \\
\midrule
General Overview & ``11 of 25 stations at Level~2; mean score 0.789. Localised but significant; immediate operational response recommended.'' \\[2pt]
Farmer & ``Immediate no-go: expect deep standing or flowing water. Do NOT attempt to access nearby fields by vehicle.'' \\[2pt]
Agronomist & ``$\Delta P = -21$\,hPa with humidity 88\% indicates frontal-system passage; prolonged soil saturation expected. Prioritise drainage assessment within 48\,h.'' \\[2pt]
Farm Manager & ``Eleven Level-2 and seven Level-1 sites require route updates; reschedule field operations away from (55.394,\,10.489) for the next 24\,h.'' \\[2pt]
Government & ``Audit entry: 11 Level-2 detections at 2026-02-15 17:09:43; actions assigned to Regional Emergency Coordinator; municipal notification pending.'' \\[2pt]
Insurance & ``Flooding score 3.003 at (55.394,\,10.489); expected total-loss for ground-floor contents; highest claim priority.'' \\
\bottomrule
\end{tabular}
\end{table}

\section{Expert Architectural Review}
\label{sec:evaluation}

This section reports a structured expert review of the architecture.
The goal is to assess the soundness of the architectural design and the novelty of the two-layer pattern from a technical perspective.

\subsection{Method}
\label{sec:eval_method}

The evaluation answers a deliberately scoped question: \emph{do the architectural claims C1 and C2 (unidirectional invariant, persona-as-configuration) hold under expert architectural review?}
It does not assess whether the generated reports are accurate, actionable, or trusted by domain stakeholders; that is left to a future study with agricultural domain experts. The panel composition (architecture, ML, edge, and security expertise) and the dimensions reflect this scope.

Where applicable, each dimension maps to ISO/IEC~25010 quality characteristics and subcharacteristics~\cite{iso_25010}:
\emph{D2}~Separation of Concerns (modifiability),
\emph{D3}~Extensibility (modifiability, flexibility),
\emph{D4}~Practical Viability (operability, reliability), and
\emph{D5}~Generalisability (flexibility, reusability).
Each dimension probes one or more drivers of \S\ref{sec:drivers}: D2 tests D-Det and D-Rep, D3 tests D-Stk and D-Sub, D4 tests D-Cost and D-Conn, and D5 tests transfer of the pattern as a whole. D1 (Architectural Novelty) is an originality check rather than a driver.

We conducted a structured walk-through session with four researchers whose complementary expertise spans the key architectural concerns of the system:
(E1)~an ML expert with experience in model compression and edge deployment;
(E2)~a security and LLM expert specialising in prompt robustness and trust boundaries;
(E3)~an edge computing expert with a background in IoT architectures; and
(E4)~a simulation and human-in-the-loop expert focusing on decision-support interfaces.

The session had three phases:
(1)~a live walk-through of the dashboard, map, and report generation;
(2)~independent ratings on the five quality dimensions above using a 5-point Likert scale (1~=~Strongly Disagree to 5~=~Strongly Agree);
(3)~a moderated open discussion of the architecture's strengths, weaknesses, and possible improvements, recorded as written notes by two of the authors and analysed through open thematic coding, yielding the four themes in the next subsection.

\subsection{Results}

Table~\ref{tab:eval_results} presents the ratings as median and range. Means are omitted because they over-state precision with $N{=}4$ on a 5-point Likert.
The highest-rated dimension was D2 (Separation of Concerns, median~5, range 4 to 5), reflecting strong agreement that the unidirectional data-flow invariant effectively isolates deterministic and non-deterministic concerns.
D3 (Extensibility, median~4.5, range 4 to 5) was also rated favourably.
D1 (Architectural Novelty, median~4, range 4 to 5) received uniformly positive ratings. E3 noted that formalising the Persona pattern as an architectural concern was the most distinctive aspect.
D4 (Practical Viability, median~3.5, range 3 to 4) received more moderate ratings, with E1 and E3 noting dependence on connectivity and operator training.
D5 (Generalisability, median~4, range 3 to 4) was positive but E4 observed that demonstrating transfer requires instantiation in at least one additional domain.
Inter-rater spread did not exceed one Likert point on any dimension, indicating consistent assessment despite the small panel.

\begin{table}
\centering
\caption{Expert panel ratings (1 to 5 Likert scale). E1: ML, E2: Security/LLM, E3: Edge, E4: Simulation/HiL.}
\label{tab:eval_results}
\begin{tabular}{@{}lcccccc@{}}
\toprule
Dimension & E1 & E2 & E3 & E4 & Median & Range \\
\midrule
D1: Architectural Novelty    & 4 & 4 & 5 & 4 & 4   & 4 to 5 \\
D2: Separation of Concerns   & 5 & 5 & 4 & 5 & 5   & 4 to 5 \\
D3: Extensibility            & 5 & 4 & 4 & 5 & 4.5 & 4 to 5 \\
D4: Practical Viability       & 3 & 4 & 3 & 4 & 3.5 & 3 to 4 \\
D5: Generalisability          & 4 & 4 & 4 & 3 & 4   & 3 to 4 \\
\bottomrule
\end{tabular}
\end{table}

\subsection{Qualitative Findings}

Thematic coding of the open discussion yielded four themes.

\textbf{Strength: Trust boundary clarity.} All panellists commended the architectural separation between the safety-critical edge path and the advisory LLM layer. E2 highlighted that the no-write-back invariant is critical given prompt-injection and hallucination risks, since the generative layer cannot corrupt edge decisions when its output is consumed by a human, not by the control plane.

\textbf{Strength: Stakeholder extensibility.} E1 and E4 noted that the prompt-configuration-only mechanism yields a favourable effort-to-value ratio, keeping the data pipeline and UI invariant while varying only the prompt template.

\textbf{Concern: Absence of stakeholder validation.} All panellists identified the lack of end-user evaluation as the most significant limitation. E4 emphasised that expert assessment does not substitute for empirical evidence about report accuracy and actionability, and E1 suggested that even a small-scale study with three to five stakeholders per role would substantially strengthen the claims.

\textbf{Concern: Real-time gap.} E3 and E1 raised the batch-mode limitation, suggesting MQTT-based streaming as a priority extension. Further observations included prompt-template version-stamping and on-device vision-language models.

\subsection{Threats to Validity}

\emph{External validity.} The architecture has been instantiated in a single domain (agricultural flood monitoring), and generalisation remains to be demonstrated.
\emph{Internal validity.} The expert panel comprised four researchers whose complementary expertise covers the system's key architectural concerns. A larger panel could surface additional issues, and panellists aware of the project goals may introduce positive bias.
\emph{Construct validity.} The five dimensions were grounded post hoc in ISO/IEC~25010 categories (\S\ref{sec:eval_method}).
\emph{Scope.} The panel validates architectural, not end-user, properties. None of the panellists are agricultural stakeholders, and ratings are not evidence about report accuracy, persona differentiation, or stakeholder trust.

\section{Discussion}
\label{sec:discussion}

\subsection{Architectural Trade-Offs}

Each trade-off corresponds to a driver of \S\ref{sec:drivers}.
\emph{Latency asymmetry} (D-Det): edge decisions are per-frame with bounded tail latency; LLM reports tolerate multi-second cloud-API latency because they serve post-hoc interpretation, not real-time hazard detection.
\emph{Cost control} (D-Cost): on-demand invocation keeps a typical 2\,000 to 3\,000-token report under one cent at GPT-5-mini~\cite{openai_gpt5mini} pricing as of Q1~2026.
\emph{Connectivity asymmetry} (D-Conn): the edge system operates entirely offline during field traversal, while the LLM layer requires connectivity only at the base station after a run. Substitutability of \texttt{llm\_report} (\S\ref{sec:llm_boundary}) provides a migration path to local models when base-station connectivity is unreliable.
\emph{Hallucination containment} (motivates \S\ref{sec:gen_reliability}): the insurance excerpt of Table~\ref{tab:report_excerpts} extrapolates beyond the schema, and the unidirectional invariant keeps such fabrications out of the control plane.

\subsection{Generative Reliability as an Architectural Concern}
\label{sec:gen_reliability}

The unidirectional invariant (\S\ref{sec:two_layer}) prevents LLM non-determinism from corrupting the control plane, but it does not prevent the more frequent failure mode in this class of systems: a confidently wrong report that a stakeholder acts on.
Hallucination, extrapolation beyond input fields, and ungrounded recommendations remain on the safety surface.
Aligned with multi-layered guardrail proposals for foundation-model-based systems~\cite{swiss_cheese_ai}, we treat generative reliability as a property of the integration boundary rather than of the LLM itself, and locate four established mitigation patterns on the architecture in Figure~\ref{fig:architecture}:

\begin{description}
\item[M1: Input grounding by construction.] \texttt{llm\_report} sees only the structured JSON record schema (Table~\ref{tab:json_fields}), so free-text context cannot shift the prompt distribution. This attaches at the data-injection step (\S\ref{sec:reporting}, step~1) and is the only mitigation already realised in the prototype.
\item[M2: Field-level attribution.] Prompt templates can require the LLM to cite the source field for each numerical claim (e.g., \emph{``combined score 3.003 $\to$ flooded, source field \texttt{scores.combined}''}). Attaches at the prompt-template seam, changing the persona contract without touching the data pipeline.
\item[M3: Schema- or grammar-constrained generation.] Reports can be generated against a per-stakeholder output schema via guided generation that constrains LLM token sampling to a regular expression or context-free grammar~\cite{willard_outlines}. Eliminates fabricated fields by construction. Attaches at the API-submission step (\S\ref{sec:reporting}, step~3).
\item[M4: Abstention on missing input.] The dashboard can refuse to invoke the LLM when required JSON fields are absent or anomalies exceed plausibility bounds, surfacing a structured ``insufficient input'' notice instead of a confident report. Attaches at the on-demand invocation seam (\S\ref{sec:llm_boundary}).
\end{description}

We do not claim to have solved hallucination, and M1--M4 are not exhaustive: retrieval-augmented grounding~\cite{gao_rag_survey,iot_llm} and runtime guardrail layering~\cite{swiss_cheese_ai} attach at the same seams. The boundary admits each mitigation as a configuration or middleware change, which is what makes the pattern transferable to other context-aware, smart cyber-physical systems with similar reliability surfaces.

\subsection{Future Work}
\label{sec:future_work}

The most significant gap, identified by the expert review (\S\ref{sec:evaluation}), is empirical validation with actual domain stakeholders.
A future study with farmers, agronomists, farm managers, government-agency representatives, and insurers would assess factual accuracy against the JSON record, actionability, and role-targeting appropriateness of the generated reports. Beyond this primary study, the following extensions are scoped briefly.

\textbf{Autonomous persona selection.} The versioned template repository (C2) is a substrate for autonomic selection~\cite{kephart_autonomic,krupitzer_self_adaptive,ibm_autonomic_blueprint}: a MAPE-K loop scoring template metadata against authentication context and recent JSON state, with fallback to the general-overview persona on low confidence. The selection policy adapts; the templates and the C1/C2 invariants do not.

\textbf{Streaming ingest.} Streaming JSON over MQTT would enable live monitoring (weakening D-Conn) and address the panel's real-time gap, provided lossy delivery does not threaten D-Rep.

\textbf{Mitigation prototypes.} Among the patterns of \S\ref{sec:gen_reliability}, M3 (schema-constrained generation~\cite{willard_outlines}) and M4 (abstention) are the obvious next prototypes: largest reliability gain for smallest implementation cost. Either would convert the seams analysis into a working demonstration.

\textbf{Multi-LLM comparison and template provenance.} A multi-LLM accuracy/cost study is the natural empirical follow-up to the stakeholder study above. Because D-Sub treats model selection as a configuration boundary, it belongs to the follow-up. Content-hashing prompt templates alongside the JSON schema would further strengthen D-Rep provenance traceability.

\section{Conclusion}
\label{sec:conclusion}

We presented \emph{persona-as-configuration}, an architectural pattern that pairs a deterministic edge plane with a generative stakeholder-reporting layer through two invariants: unidirectional consumption of structured edge decision logs (C1), and stakeholder adaptation as a versioned prompt-template artifact, not runtime improvisation (C2).
We instantiated the pattern over a previously published edge-based standing-water detection system~\cite{larsen_edge_detection} and analysed how the integration boundary admits four standard generative-reliability mitigations (M1 to M4) as configuration- or middleware-level extension points.
A structured expert review scoped to the architectural claims and tied to the ISO/IEC~25010 quality model rated the pattern favourably, strongest on separation of concerns (median~5) and extensibility (median~4.5). End-user evaluation with agricultural stakeholders is planned for future work.
The pattern targets context-aware, smart cyber-physical domains where edge devices produce structured decision logs for heterogeneous stakeholders, such as predictive maintenance, clinical decision support, and autonomous-vehicle fleet operations.
{\sloppy Source code, prompt templates, and synthetic decision-log traces are publicly available at \url{https://github.com/Oliver1703dk/generative-reporting-for-agricultural-floods}.\par}

\bibliographystyle{splncs04}
\bibliography{refs}

\end{document}